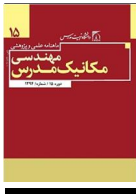
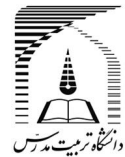

# شبیه سازی عددی شکست جت مایع با استفاده از روش هیدرودینامیک ذرات هموار (SPH)

مجید پورعبدیان[1]، پوریا امیدوار[2]*، محمدرضا مراد[3]

1- کارشناسی ارشد، مهندسی هوافضا، دانشگاه صنعتی شریف، تهران
2- استادیار، مهندسی مکانیک، دانشگاه یاسوج، یاسوج
3- استادیار، مهندسی هوافضا، دانشگاه صنعتی شریف، تهران
* یاسوج، صندوق پستی 75914-353، omidvar@yu.ac.ir

## چکیده

در این مقاله شکست جت مایع با استفاده از روش هیدرودینامیک ذرات هموار که یک روش عددی لاگرانژی بدون شبکه است، شبیه سازی شده است. بدین منظور، ابتدا معادلات حاکم بر سیال براساس روش هیدرودینامیک ذرات هموار گسسته سازی شد. در این تحقیق از کد متن باز اسفیویکس برای حل عددی استفاده و همچنین با افزودن اثرات کشش سطحی، که مذکور توسعه داده شده است. سپس روش ارائه شده با استفاده از مساله نمونه شکست سد با مانع اعتبارسنجی اولیه شد. در نهایت شبیه سازی جریان جت دو بعدی و بررسی رفتار شکست شدن آن برای جریان تک فاز انجام شده است. طول شکست مایع در رژیم ریلی برای شرایط مختلف جریان نظیر اعداد رینولدز و وبر متفاوت، محاسبه و با یک رابطه تجربی اعتبارسنجی شده است. تمامی حل های عددی برای دو تابع میانیاب وندلند و اسپلاین مکعبی انجام گرفت و در تمامی موارد تابع میانیاب وندلند منجر به نتایج با دقت بهتری شد. همچنین نتایج حاضر با روش عددی ام پی اس برای مایع غیرلزج نیز مقایسه شده است. مدل سازی های انجام شده نشان داد که روش هیدرودینامیک ذرات هموار می تواند به عنوان یک روش بهینه برای مدل سازی پدیده شکست جت مایع مورد استفاده قرار گیرد.

---

# Numerical simulation of liquid jet breakup using smoothed particle hydrodynamics (SPH)

## Majid Pourabdian[1], Pourya Omidvar[2]*, Mohammad Reza Morad[1]

1- Department of Aerospace Engineering, Sharif University of Technology, Tehran, Iran
2- Department of Mechanical Engineering, Yasouj University, Yasouj, Iran
* P.O.B. 75914-353, Yasouj, Iran, omidvar@yu.ac.ir

## ABSTRACT

In this paper, breakup of liquid jet is simulated using smoothed particle hydrodynamics (SPH) which is a meshless Lagrangian numerical method. For this aim, flow governing equations are discretized based on SPH method. In this paper, SPHysics open source code has been utilized for numerical solutions. Therefore, the mentioned code has been developed by adding the surface tension effects. The proposed method is then validated using dam break with obstacle problem. Finally, simulation of two-dimensional liquid jet flow is carried out and its breakup behavior considering one-phase flow is investigated. Length of liquid breakup in Rayleigh regime is calculated for various flow conditions such as different Reynolds and Weber numbers and the results are validated by an experimental correlation. The whole numerical solutions are accomplished for both Wendland and cubic spline kernel functions and Wendland kernel function gave more accurate results. The results are compared to MPS method for inviscid liquid as well. The accomplished modeling showed that smoothed particle hydrodynamics (SPH) is an efficient method for simulation of liquid jet breakup phenomena.

## 1- مقدمه

شکسته شدن مایع پیوسته به قطرات ریز یا اصطلاحا فرآیندهای شکست مایع و اتمیزه شدن مایع در بسیاری از کاربردهای مهندسی نظیر افشانه[1] سوخت در محفظه احتراق موتورهای جت و احتراق داخلی ظاهر می شود. به صورتی که اندازه افشانه سوخت بر بازده موتور بسزایی دارد [1]. یکی از

موضوعات جالب در زمینه مهندسی-پزشکی، استفاده از انژکتورهای نانومتر و میکروجت برای کنترل داروهای وارده به بدن است که طول شکست مایع ارتباط مستقیم با عمق نفوذ دارو در بدن بیمار دارد [3,2]. دینامیک جت مایع طیف گسترده ای از خواص فیزیکی را شامل می شود که از این بین می توان به کشش سطحی مایع، لزجت و چگالی مایع در مقایسه با محیط اطراف آن اشاره کرد. در کاربردهای کوچکتر نظیر نانومتر، جت با نوسانات دمایی

<hr>

1- Spray

حساس هستند. در حالی که در کاربردهای بزرگ، فعل و انفعالات گرانشی اهمیت می یابند. پیچیدگی عوامل فیزیکی، تحلیل پدیده شکست جت مایع را سخت می کند. در شرایط ایده آل که می توان تاثیرات برخی از عوامل فیزیکی را صرفنظر کرد، حل تحلیلی این پدیده امکان پذیر است. در سال های اخیر شکست جت مایع به دلیل کاربرد فراوان آن در انژکتور محفظه احتراق موتورهای مختلف و افزایش بازدهی موتور با ایجاد افشانه بهینه و همچنین کاربرد آن در مسائل بیولوژیک مورد توجه قرار گرفته است.

طول شکست جت به وسیله ریکو و اسوالدینگ [4] برای جریان های جت آشفته، تاناساوا و تویودا [5] برای جریان های جت آرام و آشفته و پارک و همکاران [6] برای نفوذ جت به همراه تبخیر مطالعه شده است. برای مطالعه رفتار شکست جت به صورت عددی به یک روشی نیازمندیم که توانایی تغییر شکل بزرگ در سطوح آزاد را داشته باشد. لذا اکثر مطالعات انجام گرفته در این زمینه به صورت تجربی و تحلیلی بوده است [7-12]. همچنین بررسی و شبیه سازی این پدیده با روش های مبتنی بر شبکه یا امکان پذیر نیست و یا نیازمند محاسبات عددی بسیار زمان بر و سنگین است [14و13]. ریچاردز و همکاران شکست یک جت مایع-مایع را با استفاده از روش حجم سیال[2] و روش سی اس اف[3] مدلسازی و با نتایج تجربی مقایسه کردند [15].

در سال های اخیر، شیباتا و همکاران [16] شکست جت مایع غیرلزج را که از یک نازل خارج می شد، با استفاده از روش ام پی اس[5] مطالعه کردند و نتایج خود را برای توزیع اندازه قطرات تشکیل شده گسترش دادند. گائزنمولر و همکاران در سال 2007 از روش اس پی اچ[5] برای مدلسازی پاشش دیزل استفاده کردند [17]. در سال 2012، تاکاشیما و همکاران [18] شکست جت مایع خروجی از یک نازل سیلندری را با استفاده از روش اس پی اچ تراکم ناپذیر[6] در حالت پالسی مطالعه کردند. سبروتکین و یوج مدلسازی شکست جت مایع بسیار لزج را با استفاده از روش اس پی اچ تصحیح شده انجام دادند و عدد وبر بحرانی برای انتقال جریان از جت به حالت چکه کردن[7] را بررسی کردند [19]. همچنین در همین سال جریان دوفازی جت با در نظر گرفتن چگالی های یکسان و به صورت محدود توسط هوفلر و همکاران با روش اس پی اچ شبیه سازی شد [20]. هوفلر و همکاران در سال های 2010 و 2013 هم شبیه سازی بر روی جریان جت مایع که از یک مخزن بر اثر نیروی گرانش تخلیه می شد را با روش اس پی اچ انجام دادند و به بررسی تاثیرات کشش سطحی بر میدان سرعت جریان پرداختند [22و21].

با توجه به تحقیقات پیشین بیان شده، جریان شکست جت مایع و به ویژه طول شکست آن بسیار محدود با استفاده از روش اس پی اچ مورد مطالعه قرار گرفته است و تحلیل این پدیده با استفاده از روش اس پی اچ نیازمند پژوهش و تحقیقات بیش تری است. در این مقاله برای نخستین بار طول شکست جت مایع خروجی از یک مخزن در رژیم رِیلی[8] با استفاده از معادلات استاندارد اس پی اچ برای عدد رینولدزهای کوچکتر از 1200 مورد مطالعه قرار گرفته است. همچنین تاثیر جت میانیاب در این پدیده که تاکنون مورد مطالعه قرار نگرفته، در این تحقیق ارزیابی و بررسی شده است. برای

---

مدل سازی عددی این پدیده از یک کد متن باز اسفیزیکس[9] استفاده شده است [23]. این کد برای حل عددی جریان های سطح آزاد و مسائل هیدرودینامیکی ارائه شده و دارای مسائل نمونه محدودی است، لذا برای مدل سازی عددی جریان شکست جت مایع، در این تحقیق کد متن باز اسفیزیکس با افزودن اثرات کشش سطحی توسعه داده شده است. در ادامه به توضیح اصول کلی روش اس پی اچ پرداخته شده است و در بخش سوم نتایج حاصل شده ارائه شده و مورد بررسی قرار گرفته است.

## 2- حل عددی به روش هیدرودینامیک ذرات هموار

### 2-1- هیدرودینامیک ذرات هموار

روش اس پی اچ اولین بار توسط گینگلد و موناگان و به طور جداگانه توسط لوسی در سال 1977 معرفی شد [24]. از آنجا که اولین نسخه اس پی اچ اصل پایستگی مومنتوم خطی و زاویه ای را ارضا نمی کرد، درسال 1982 گینگلد و موناگان الگوریتم اولیه آن را با استفاده از لاگرانژین متناظر ذرات بهبود بخشیدند تا برای سیال تراکم پذیر فاقد استهلاک، پایستگی مومنتوم خطی و زاویه ای را ارضا شود [25]. موناگان برای اولین بار با استفاده از فرض تراکم پذیری مصنوعی[10]، روش اس پی اچ را برای حل جریان سیال تراکم ناپذیر بکار برد [26]. در سال 1998 کومینز و رودمن با بکاربردن روش تصویرسازی[11] جریان سیال در یک حفره را شبیه سازی کرده و اولین باری بود که برای تصویر سازی برای جریان به کمک روش اس پی اچ بکار برده می شد [27]. در سال 2003 کلگروسی و لندرینی با تغییر در معادله حالت توانستند سیالات چند فاز با اختلاف چگالی بالا را شبیه سازی کنند [28]. در سال 2006 هو و آدامز روشی را برای مدلسازی کشش سطحی ارائه کردند [29] و سپس در سال 2007 هو و آدامز به مدلسازی جریان سیال چند فازی پرداختند [30]. در سال های 2012 و 2013 امیدوار و همکاران با توسعه روش اس پی اچ به کمک توزیع متغیر جرم ذره، اجسام شناور روی آب را شبیه سازی کردند [31،32]. با توجه به اینکه روش اس پی اچ در مدلسازی سیال چند فاز با اختلاف چگالی بالا دارای خطای زیاد است، در سال 2013 موناگان و رفیعی توانستند الگوریتمی ارائه دهند که در عین سادگی، در پایداری حل عددی جریان با نسبت چگالی بالا بسیار موثر میباشد [33]. در سال 2015 نیز امیدوار و همکاران انتشار امواج آب را درون یک کانال با استفاده از روش ترکیبی اس پی اچ و ای ال ای[12] شبیه سازی کردند و نشان دادند که این روش ترکیبی حل عددی را پایدارتر می کند و سبب کاهش نویزهای میدان فشار می شود [34]. همانطور که دیده می شود روش هیدرودینامیک ذرات هموار در مقایسه با روش های عددی پیشین جوان تر و نیازمند توسعه بیش تری است و همچنین مسائل بسیاری از طبیعت لاگرانژی دارند هنوز با این روش شبیه سازی نشده اند.

### 2-2- معادلات حاکم

معادلات اصلی حاکم بر سیال در مسائل مکانیک سیالات شامل معادلات بقای جرم (پیوستگی) و بقای اندازه حرکت (ممنتوم) می باشد (1و2):

$$\frac{d\rho}{dt} = -\rho \nabla \cdot u \tag{1}$$

$$\frac{du}{dt} = -\frac{1}{\rho}\nabla P + g + \theta + F_s \tag{2}$$

---

در معادلات بالا $\frac{d}{dt}$ معرف مشتق مادی، $\rho$ جرم مخصوص، $u$ بردار سرعت، $P$ فشار، $g$ شتاب گرانشی، $\Theta$ و $F_s$ به ترتیب معرف عبارت های پخش[1] و کشش سطحی می باشند. در فضای فیزیکی، می توان سیال را توسط تعداد محدودی از حجم های ماکروسکوپیک توصیف کرد. در روش اس پی اچ یک ذره $a$ نماینده یک حجم ماکروسکوپیک از سیال است. هر ذره از سیال، حامل اطلاعات مربوط به جرم، چگالی، فشار، سرعت، موقعیت و دیگر کمیت های مربوط به ماهیت جریان و سیال می باشد. جرم در تمام مراحل شبیه سازی ثابت بوده اما فشار، سرعت، موقعیت و دیگر کمیت های فیزیکی در هر گام زمانی به روز می شوند.

شالوده روش اس پی اچ بر دو گام بیان انتگرالی[2] و تقریب ذره‌ای[3] برای محاسبه متغیرهای موجود در میدان حل استوار است. به طور کلی یک متغیر و یا تابع در روش اس پی اچ در گام اول به میان یابی انتگرالی بیان می شود و سپس انتگرال مورد نظر به جمع گسسته تقریب زده می شود.

$$\phi(r) = \int_\Omega \phi(r')W(r - r', h)dr' \qquad (3)$$

در روابط روش اس پی اچ، مقدار متغیر $\Phi$ در نقطه‌ی $r$ به مختصات $\vec{r} = (x, z)$ به صورت انتگرال حجم بر روی ناحیه پشتیبانی $\Omega$ می باشد (رابطه (3)). $W$ اصطلاحا تابع میانیاب[4] خوانده می شود. $h$ طول هموارسازی[5] است که ناحیه‌ی تاثیر اطراف یک ذره را مشخص می‌کند و با افزایش آن، ناحیه تاثیرگذار در میان یابی تابع $\Phi$ بزرگ تر می شود. مقدار طول هموارسازی $h$ در شبیه‌سازی‌های مختلف براساس فاصله اولیه ذرات همراه با یک ضریب بیان می شود که انتخاب این طول معمولاً بستگی به نوع مسئله دارد. این طول در تمامی شبیه سازی های انجام شده در این مقاله $h = 1.3\,\Delta x$ در نظر گرفته شده است که $\Delta x$ فاصله اولیه بین ذرات می‌باشد. توابع میانیاب، توابع یکه هستند، به این معنا که انتگرال روی حجم آنها برابر یک است و در نتیجه توابع ثابت، به طور دقیق میان یابی می شوند. انتخاب نوع تابع میانیاب در موفقیت این روش تاثیر بسزایی دارد. در هر روش جدیدی، معادلات حاکم باید به نحوی گسسته شوند تا بتوان معادلات دیفرانسیل حاکم بر سیستم را با مجموعه‌ای از معادلات جبری تقریب زد. در روش اس پی اچ این گسسته سازی در گام دوم با تقریب انتگرال رابطه (3) انجام می شود.

$$\phi(r) \approx \sum_b \phi_b \frac{m_b}{\rho_b} W_{ab} \qquad (4)$$

که در رابطه (4)، $a$ مشخص کننده ذره مرکزی، $b$ بیانگر ذرات مجاور ذره‌ی مرکزی، $m$ جرم، $\rho$ جرم مخصوص و $W_{ab} = W(r_a - r_b, h)$ بیانگر تابع درون‌یابی نسبت به ذره‌ی $a$ می‌باشد. نسبت $\frac{m_b}{\rho_b}$ جایگزین المان حجم $dr'$ در بیان انتگرالی شده است. در شکل 1 بیان تعریف ذره را در مقایسه با ذره جایگذاری شده مشاهده کرد.

یکی از مزایای میانیابی توابع به روش اس پی اچ این است که مشتقات یک تابع به صورت تحلیلی به دست می آیند که در مقایسه با روشی مانند اختلاف محدود که از فاصله موجود بین نقاط همسایه استفاده می شود، ساده‌تر است [35]. در این روش گرادیان متغیر $\Phi$ را می‌توان بصورت تابعی از $\Phi$ و مشتقات تابع میانیاب که در ادامه آمده، تعریف کرد.

$$\nabla\phi(r_a) \approx \sum_b \frac{m_b}{\rho_b}(\phi_b - \phi_a)\nabla_a W(r_{ab}) \qquad (5)$$

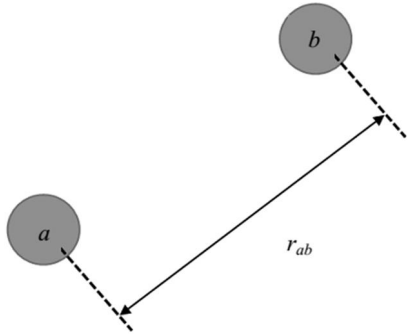

**Fig. 1** Definition of particle $a$ in the neighborhood of particle $b$ in distance $r_{ab}$

**شکل 1** تعریف ذره $a$ در مجاورت ذره $b$ در فاصله $r_{ab}$

در رابطه (5) کمیت $\nabla_a W(r_{ab})$، معرف گرادیان تابع میانیابی بوده که در مرکز موقعیت ذره‌ی a محاسبه شده است. گسسته سازی و تخمین معادله چگالی در روش اس پی اچ از اهمیت بسیار زیادی برخوردار است چرا که اصولا چگالی توزیع ذرات و طول هموارسازی را معین می کند. محاسبه چگالی در روش اس پی اچ به طور معمول به دو طریق انجام می‌گیرد. در روش اول می‌توان چگالی را بطور مستقیم از فرمولاسیون درون‌یابی روش اس پی اچ که در رابطه (4) بیان شد، به صورت زیر محاسبه کرد.

$$\rho_a \approx \sum_b m_b W_{ab} \qquad (6)$$

اما معادله (6) به این دلیل که باید بار چگالی و در گام دوم سرعت را محاسبه کرد، به یک حلقه اضافی روی همه‌ی ذرات نیاز دارد [19]. یک رابطه کاراتر برای حل عددی معادله پیوستگی، استفاده از فرم لاگرانژی آن (1) می باشد. در نتیجه می‌توان با حل معادله‌ی پیوستگی و رابطه‌ی دیورژانس[6] روش اس پی اچ، مقدار چگالی ذرات را محاسبه کرد.

$$\frac{d\rho_a}{dt} = \sum_b m_b\, u_{ab} \cdot \nabla_a W_{ab} \qquad (7)$$

در رابطه (7) متغیر $u_{ab}$ معرف بردار سرعت اختلاف سرعت بین $a$ و $b$ می‌باشد. از دیدگاه جریان تراکم پذیر، به دلیل ثابت بودن تعداد کل ذرات، جرم کل سیستم بر اساس روابط (6) و (7) ابقاء می شود. از دیدگاه جریان تراکم ناپذیر، دیورژانس سرعت به خودی خود صفر نمی شود و در نتیجه هر دو معادله پیوستگی مذکور به طور دقیق پایسته نیستند [36]. همچنین به دلیل اینکه در معادله بقای ممتنوم باید گرادیان تابع میانیاب محاسبه شود، استفاده از رابطه (7) انتخاب بهتری است چرا که می توان گرادیان تابع را در یک زیرروال[6] محاسبه کرد.

بدست آوردن معادله ممنتوم به روش اس پی اچ شبیه گسسته سازی معادله پیوستگی می باشد. عبارت گرادیان فشار در رابطه (2) را می‌توان به طور معمول برحسب فرمولاسیون روش اس پی اچ بصورت زیر محاسبه کرد.

$$\frac{1}{\rho_a}\nabla P_a \approx \sum_b m_b\left(\frac{P_a}{\rho_a^2} + \frac{P_b}{\rho_b^2}\right) \cdot \nabla_a W(r_{ab}) \qquad (8)$$

برای در نظرگرفتن تنش لزجتی و عبارت پخش در معادله ممتنوم می‌توان دو راهکار را مدنظر قرار داد. در راهکار اول می توان از تنش های لزجتی جریان آرام در معادله ممتنوم استفاده کرد. اما در راهکار دوم می توان با مدلسازی تنش های لزجتی با ترمی به نام لزجت مصنوعی[7] معادله ممتنوم را حل کرد. عبارت لزجت مصنوعی معمولا به ترم فشار فیزیکی اضافه

---

6- Subroutine
7- Artificial Viscosity

---

---

1- Diffusion Terms
2- Integral Representation
3- Particle Approximation
4- Kernel Function
5- Smoothing Length

عاملی برای کاهش مساحت سطح کل است (و متعاقبا انرژی سطحی). برای $\delta_s$ در رابطه (11) انتخاب های متفاوتی وجود دارد. اما در هر حال این نرمال کننده باید طوری برگزیده شود که انتگرال آن روی سطح تماس برابر با واحد شود. این شرط بدین منظور است که با افزایش دقت حل، فیزیک سطح تماس به درستی بازیابی شود. در نتیجه می توان از تابع دلتای سطحی به صورت زیر استفاده کرد.

$$\delta_s = |n| \tag{13}$$

که $|n|$ اندازه بردار عمود بر سطح تماس است. سطح تماس هوا و آب می تواند توسط یک تابع به نام تابع رنگ[5] تشخیص داده شود. تابع رنگ برای محاسبه بردار نرمال و متعاقبا محاسبه انحنای سطح تماس بکار گرفته می شود. در بیان تقریب ذرهای روش اس پی اچ تابع رنگ به صورت زیر بیان می شود:

$$C_a = \sum \frac{m_b}{\rho_b} W_{ab} \tag{14}$$

قبل از شروع حل عددی، تابع رنگ برای ذرات آب یک و برای ذرات هوا صفر مفروض می شود. در نتیجه در طول حل عددی در نواحی دور از سطح تماس آب و هوا، تابع رنگ برای ذرات آب و هوا تقریبا ثابت و به ترتیب برابر یک و صفر باقی می ماند. در حالی که در نزدیکی های سطح آزاد و سطح تماس به سیال تابع رنگ حدود $0.4 \sim 0.5$ افت می کند. می توان بردار نرمال سطح را بر حسب گرادیان تابع رنگ به صورت زیر تخمین زد.

$$n = \nabla C \tag{15}$$

همچنین بردار یکه نرمال هم در ادامه محاسبه می شود.

$$\hat{n} = \frac{n}{|n|} \tag{16}$$

انحنای سطح تماس به صورت دیورژانس بردار یکه نرمال تعریف می‌شود.

$$\kappa = \nabla \cdot \hat{n} \tag{17}$$

بنابراین بر اساس رابطه های (15) و (17) می توان تقریب عددی بردار نرمال و انحنا را به صورت زیر نوشت [40]:

$$n_a = \sum_b \frac{m_b}{\rho_b} (C_b - C_a) \nabla_a W_{ab} \tag{18}$$

و

$$\kappa = (\nabla \cdot \hat{n})_a = \sum_b \frac{m_b}{\rho_b} (\hat{n}_b - \hat{n}_a) \cdot \nabla_a W_{ab} \tag{19}$$

در نتیجه بر طبق این فرمول بندی، سطح تماس منطقه ای است که مقدار تابع رنگ تغییر زیادی کند. سطح تماس دارای عرض محدودی می باشد. هنگامی که تعداد ذرات به سمت بی نهایت میل کند، عرض سطح تماس نیز به سمت صفر می گراید. همانطور که اشاره شد در نواحی دور از سطح تماس، تابع رنگ تقریبا ثابت است و در نتیجه هنگامی که رابطه (16) برای ناحیه خارج از سطح تماس استفاده شود، بردارهای $n$ بسیار کوچک می شوند و این موضوع منجر به مقادیر بزرگ $\hat{n}$ و در نتیجه $\kappa$ می شود. این اتفاق باعث می شود تا جهتها و مقادیر نیروی کشش سطحی به درستی محاسبه نشود. یکی از راهها برای جبران این نادرستی، فیلتر کردن $\hat{n}$ به شکل زیر است.

$$N_a = \begin{cases} 1 & , & \text{اگر } |n| > \varepsilon \\ 0 & , & \text{در غیر اینصورت} \end{cases} \tag{20}$$

و

---

می شود و باعث می شود که متغیرهای جت هموار در جریان نفوذ کرده و انرژی ترم های فرکانس بالا حذف شود. برای وارد کردن تاثیرات لزجت در معادله ممنتوم به روش اس پی اچ، مونلاگان عبارتی تحت عنوان لزجت مصنوعی $\Pi_{ab}$ را به معادله ممنتوم به صورت زیر اضافه کرد [37]:

$$\frac{du_a}{dt} = -\sum_b m_b \left( \frac{P_a}{\rho_a^2} + \frac{P_b}{\rho_b^2} + \Pi_{ab} \right) \cdot \nabla_a W(r_{ab}) \tag{9}$$

که در آن $\frac{dr_a}{dt} = u_a$ در واقع ترم دیفیوژن $\Theta$ بیان شده در معادله (2) با ترم لزجت مصنوعی $\Pi_{ab}$ در گرادیان فشار همسان سازی شده است. ویسکوزیته مصنوعی سبب افزایش پایداری حل عددی شده که به صورت زیر شبیه‌سازی می‌شود:

$$\Pi_{ab} = \begin{cases} -\frac{\alpha_{vis} h \overline{c_{ab}}}{\overline{\rho_{ab}}} \frac{u_{ab} \cdot r_{ab}}{r_{ab}^2 + \eta^2} & , & u_{ab} \cdot r_{ab} < 0 \\ 0 & , & u_{ab} \cdot r_{ab} \geq 0 \end{cases} \tag{10}$$

$\alpha_{vis}$ پارامتر ثابتی است که در مسائل مختلف می تواند تغییر کند و معمولا در مسائل هیدرودینامیکی بین $[0.01-1]$ فرض می شود [38]. $\overline{c_{ab}} = \frac{c_a + c_b}{2}$ و $\overline{\rho_{ab}} = \frac{\rho_a + \rho_b}{2}$ به ترتیب چگالی و سرعت صوت متوسط است و $c_a$ و $c_b$ به ترتیب سرعت های صوت ذره $a$ و $b$ است. $h$ طول هموارساز و همچنین $\eta^2 = 0.01 h^2$. پارامتر $\eta$ سبب جلوگیری از واگیری حل عددی در شرایطی که دو ذره به هم نزدیک می شوند، میگردد.

## 2-3- کشش سطحی

همانطور که قبلا بدان اشاره شد، در این مقاله حل عددی از کد متن باز اسفیزیکس استفاده شده است [23]. به دلیل اینکه این کد متن باز از اثرات کشش سطحی صرفنظر کرده، در نتیجه در این تحقیق این کد با افزودن اثرات کشش سطحی توسعه داده شده است. برای مدل سازی اثرات کشش سطحی در قالب روش اس پی اچ تلاش هایی صورت گرفته است که اکثر آنها از روش نیروی سطحی پیوسته[1] استفاده کرده اند [39]. در این روش کشش سطحی به عنوان یک نیروی حجمی در یک ناحیه نازک در نزدیکی سطح تماس دو سیال در نظر گرفته می‌شود. این روش این امکان را می دهد تا نیازی به اعمال محدودیت در هندسه سطح جریان نباشد. موریس [40] بر اساس همین تئوری نیروی سطحی پیوسته، روشی را برای مدلسازی کشش سطحی با استفاده از انحنا[2]ی سطح تماس و تابع دلتای سطحی[3] ارائه کرد. این متغیرها توسط یک میدان رنگ[4] محاسبه می شوند.

نیروی کشش سطحی به صورت عمودی بر سطح تماس دو سیال اعمال می شود و انرژی سطحی را کاهش می دهد. بر طبق تئوری سی اس اف کشش سطحی $F_s$ به یک نیرو بر واحد حجم تبدیل می شود.

$$F_s = f_s \delta_s \tag{11}$$

در رابطه (11)، $\delta_s$ یک تابع نرمال کننده (تابع دلتای سطحی) است و $f_s$ نیرو بر واحد سطح به صورت زیر می باشد.

$$f_s = \sigma \kappa \hat{n} \tag{12}$$

که در رابطه بالا $\sigma$ ضریب کشش سطحی، $\hat{n}$ بردار یکه عمود بر سطح تماس و $\kappa$ انحنای سطح تماس هستند. در این روش فرض شده است که کشش سطحی در تمام سیال ثابت و از گرادیان سطح چشم پوشی شده است. نیرو بیان شده در رابطه (12) به صورت عمود بر سطح تماس و بر انحنای محلی سطح عمل میکند. این نیرو، در نواحی دارای انحنای زیاد را همواره می کند و

---

$$\hat{n}_a = \begin{cases} \dfrac{n_a}{|n_a|} & , \quad \text{اگر } N_a = 1 \\ 0 & , \quad \text{در غیر اینصورت} \end{cases} \tag{21}$$

در رابطه (20) معمولاً $\varepsilon$ برابر $0.01/h$ در نظر گرفته می‌شود. بنابراین رابطه (19) به صورت زیر اصلاح می‌شود [40].

$$(\nabla \cdot \hat{n})_a^* = \sum_b \min(N_a, N_b) \frac{m_b}{\rho_b} (\hat{n}_b - \hat{n}_a) \cdot \nabla_a W_{ab} \tag{22}$$

همچنین می توان با نرمال کردن انحنا به جواب های دقیق تری نیز رسید. رابطه (23) تخمینی مطمئن از انحنای سطح تماس می دهد.

$$\kappa = (\nabla \cdot \hat{n})_a = \frac{\sum_b \min(N_a, N_b) \frac{m_b}{\rho_b} (\hat{n}_b - \hat{n}_a) \cdot \nabla_a W_{ab}}{\sum_b \min(N_a, N_b) \frac{m_b}{\rho_b} W_{ab}} \tag{23}$$

در نتیجه رابطه نهایی برای شتاب ناشی از نیروهای کشش سطحی $a_s$ و معادله نهایی ممنتوم به صورت زیر نوشته می‌شوند:

$$(a_s)_a = -\frac{\sigma_b}{\rho_a} (\nabla \cdot \hat{n})_a n_a \tag{24}$$

$$\frac{du_a}{dt} = -\sum_b m_b \left( \frac{P_a}{\rho_a^2} + \frac{P_b}{\rho_b^2} + \Pi_{ab} \right) \cdot \nabla_a W(r_{ab}) + (a_s)_a + g \tag{25}$$

## 2-4- معادله حالت

نحوه اعمال تراکم ناپذیری در روش های عددی مورد بحث و چالش بوده است. دلیل این امر را می توان نبود معادله جداگانه برای فشار در سیال های تراکم ناپذیر دانست. اگر بخواهیم برای شبیه سازی جریان های تراکم ناپذیر مانند جریان آب، همچنین از معادله حالت استفاده کنیم، اندازه قدم‌های زمانی برای ارضاء شرط کورانت یا سی اف ال[1] بسیار کوچک می شود که عملاً شبیه سازی را غیرممکن می‌سازد. برای رفع این مشکل از یک معادله حالت مصنوعی استفاده می‌شود. در این معادله سرعت صوت باید به اندازه کافی پایین باشد تا محاسبات بر مبنای آن امکان پذیر باشد و به اندازه کافی بالا باشد تا شرط تراکم ناپذیری خدشه دار نشود؛ تغییرات چگالی (یک درصد) فراتر نرود. هدف صوت مصنوعی تولید مشتق زمانی فشار است. به جای حل معادله پواسون[2] برای جریان تراکم ناپذیر در هر گام زمانی، از معادله حالت تیت[41] در این الگوریتم استفاده می شود:

$$P = B \left[ \left( \frac{\rho}{\rho_o} \right)^\gamma - 1 \right] \tag{26}$$

مقدار $\gamma$ برای مایع معمولاً 7 و برای هوا 1.4 در نظر گرفته می‌شود. متغیر $\rho_0$ چگالی مرجع و $B$ ضریب ثابتی است که به صورت $\frac{\rho_0 c_0^2}{\gamma}$ اختیار می‌شود که در آن $c_0$ سرعت صوت عددی می‌باشد.

## 2-5- تابع میانیاب

کارایی روش اس پی اچ به انتخاب تابع میانیاب بسیار وابسته است چرا که نحوه تقریب متغیرها به رفتار میانیاب بستگی دارد. در این تحقیق از دو تابع میانیاب اسپلاین مکعبی[3] و وندلند[4] استفاده شده است. تابع میانیاب اسپلاین مکعبی یک چندجمله ای به صورت زیر می باشد [42]:

$$W(r, h) = \frac{\kappa}{h^\nu} \times \begin{cases} \left(1 - \dfrac{3}{2} q^2 + \dfrac{3}{4} q^3 \right) & , \quad q \leq 1 \\ \dfrac{1}{4}(2 - q)^3 & , \quad 1 < q < 2 \\ 0 & , \quad q \geq 2 \end{cases} \tag{27}$$

که $q = \frac{r}{h}$ و مقدار $\kappa$ برای محاسبات یک‌بعدی، دوبعدی وسه‌بعدی به ترتیب $\frac{2}{3}$ و $\frac{10}{7\pi}$ و $\frac{1}{\pi}$ می‌باشد. این میانیاب تاکنون متداول ترین تابع در مقالات و شبیه سازی های انجام شده به روش اس پی اچ بوده است چرا که شبیه تابع میانیاب گوسین می باشد با این تفاوت که دارای ناحیه پشتیبانی کوچکتری است و از نظر حجم محاسباتی مزیت بهتری نسبت به میانیاب گوسین دارد.

این تابع توسط وندلند در سال 1995 ارائه شده است [43] و بصورت زیر نوشته می‌شود:

$$W(r, h) = \frac{\kappa}{h^\nu} \times \begin{cases} (1 + 2q)(2 - q)^4 & , \quad 0 \leq q < 2 \\ 0 & , \quad q \geq 2 \end{cases} \tag{28}$$

که مقدار $\kappa$ برای محاسبات یک بعدی، دوبعدی وسه‌بعدی به ترتیب $\frac{3}{4}$ و $\frac{7}{8\pi}$ و $\frac{7}{4\pi}$ می‌باشد. تابع میانیاب مرتبه‌ی پنجم مانند وندلند به شبیه سازی های عدد رینولدز خیلی پایین موجب بقای پایداری شده در حالی که تابع میانیاب کیوبیک نویزهایی را در نتایج سرعت و فشار موجب می‌شود. تابع میانیاب بهینه برای داشتن دقت و زمان حل مناسب تابع وندلند می‌باشد.

## 2-6- شرایط مرزی

اعمال شرایط مرزی در روش اس پی اچ هنوز به تکاملی مشابه اعمال شرایط مرزی در روش‌های مبتنی بر شبکه نرسیده است و احتیاج به تحقیقات بیشتری دارد. برای مثال هنوز نحوه اعمال شرط مرزی جریان ورودی و خروجی به طور کامل مشخص نیست. در این مقاله از روش نیروی دافعه[5] بین مولکولی، ذراتی که به عنوان مرز شناخته می‌شوند نیروهای مرکزی بین ذرات سیال اعمال می‌کنند. این روش در سال 1999 توسط موناگان و کس [44] با بکارگیری یک فرآیند تقریب زنی بهبود یافت. در سال 2008 راجرز و همکاران [45] برای آسان کردن شرایط مرزی پیچیده و اندرکنش بین ذرات و مرز، این روش را اصلاح کردند. در این روش نیرویی بر ذرات سیال در جهت نرمال سطح وارد می‌شود که مقدار این نیرو برابر است با:

$$f = nR(\psi)P(\xi)\varepsilon(z_t, u_\perp) \tag{29}$$

که در رابطه (28) $n$ بیانگر نرمال واحد و $R(\psi)$ تابع نیروی دافعه برحسب فاصله‌ی $\psi$ است که $\psi$ خود فاصله‌ی عمودی از دیواره‌ی مرز می‌باشد. $P(\xi)$ بیانگر این مورد است که وقتی ذرات سیال در بین دو ذره‌ی مرزی حرکت کند نیرویی که از طرف ذرات مرزی وارد می شود به مقدار ثابت و به موازات ذرات مرزی است. $\xi$، تصویر فاصله‌ی قطری ذره بر روی دو ذره‌ی مرز است. در نهایت $\varepsilon(z_t, u_\perp)$ عبارتی است که برای اصلاح روش اولیه پیشنهاد شده توسط موناگان و کس به رابطه نیروی دافعه افزوده شده است که مقدار نیروی وارده را بر حسب عمق محلی سیال و سرعت عمود بر مرز ذرات سیال، تنظیم می‌کند.

## 2-7- الگوی جستجوی ذرات

حل مسائل با تعداد ذرات زیاد با استفاده از یک الگوریتم جستجوی معمولی، مدت زمان اجرای برنامه را افزایش خواهد داد. در این تحقیق از الگوریتم جستجوی لیست اتصال[6] استفاده شده است [46]. در این روش مدت زمان برای جستجوی ذرات متناسب با $N \log N$ می‌باشد. استفاده از این الگوریتم جستجو نه‌تنها موجب کاهش مدت زمان لازم برای پردازش شده بلکه حجم

---

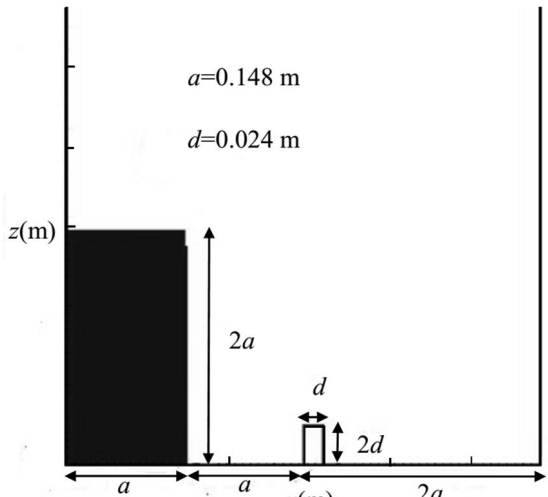

$a=0.148$ m

$d=0.024$ m

$z$(m)

$2a$

$d$

$2d$

$a$ | $a$ | $2a$

$x$(m)

**Fig. 2** Geometry of dam-break with obstacle problem

**شکل 2** هندسه مساله شکست سد با مانع

جریان شکست سد با مانع را با استفاده از دو تابع میانیاب وندلند و اسپلاین مکعبی مشاهده کرد، به صورتی که تطابق نسبتا خوبی بین حل عددی به وسیله‌ی تابع میانیاب های وندلند و اسپلاین وجود دارد. این تطابق تا زمان 0.3 ثانیه بسیار زیاد است اما مشهودترین تفاوت در نیمه دوم زمانی (0.6 - 0.3) ثانیه جداشدن بیشتر ذرات آب در ارتفاع های بالای سد برای تابع میانیاب وندلند است. همانطور که در شکل 3 مشاهده می شود، در زمان 0.1 ثانیه لایه نازکی از آب به مانع جلوی سد رسیده است. در زمان 0.2 ثانیه آب سد با مانع روبرو شده و سمت چپ بالایی و در گوشه بالایی زبانه انحنایی از مایع تشکیل می‌شود. در زمان 0.3 ثانیه زبانه و انحنای آب به حرکت خود ادامه داده و انحنای بیشتری به خود می‌گیرد. در حل های عددی با تابع میانیاب وندلند و همچنین در حل قاسمی دیده می‌شود که ذراتی از آب از انتهای انحنا جدا شده‌اند. در حالی که تابع میانیاب اسپلاین نتوانسته این جدایش را مدل کند. در زمان 0.4 ثانیه بخشی از آب با دیواره روبرویی (سمت راست) برخورد کرده و شروع به پایین آمدن می‌کند و به کف ظرف برخورد می‌کند. همچنین در زمان 0.5 ثانیه قطرات جدا شده بیشتری در انتهای بالایی انحنای آب در نتایج قاسمی و تابع وندلند دیده می‌شود. به دلیل صرفنظر کردن از وجود هوا در خارج از جریان اصلی و تاثیر مقاومت سیال هوا، نمی‌توان اندازه حباب‌های تشکیل شده در داخل آب را به طور دقیق شبیه سازی کرد. همانطور که در زمان 0.6 ثانیه مشهود است، حباب ایجاد شده درون جریان آب در دو روش عددی از نظر اندازه با هم تفاوت دارند که می توان این اختلاف را ناشی از استفاده شرط مرزی سطح آزاد توسط قاسمی دانست. در نهایت می‌توان گفت که شبیه سازی جریان تک فازی شکست سد با مانع نتایج قابل قبولی را می‌دهد اما استفاده از تابع میانیاب وندلند منجر به حصول نتایج دقیق‌تری می شود.

### 3-2- شکست جریان جت مایع

در شبیه سازی جت از یک مخزن پر شده، که از ذرات پر شده، به عنوان نازل استفاده شده است. شکل 4 هندسه نازل و ابعاد آن و همچنین قطر اریفیس[4] را نشان می دهد. لازم به توضیح است که شکل بزرگ شده در شکل 4 یک نمای نمونه و نه واقعی برای بهتر نشان دادن اریفیس نازل و ذرات اطراف آن

---
4- Orifice

حافظه‌ی جانبی مورد نیاز برای ذخیره ماتریس‌ها را نیز کاهش می‌دهد.

### 2-8- تعیین گام زمانی

مادامی که گام زمانی به اندازه کافی کوچک باشد تا پایداری و دقت حل عددی حفظ شود، نتایج حاصل از روش اس پی اچ وابستگی زیادی به انتخاب الگوریتم زمانی ندارد [47]. در این مقاله از الگوریتم پیش‌بینی- تصحیح[1] معادلات [48] استفاده شده است. تعیین گام زمانی در روش‌های مبتنی بر ذره[2] یک پارامتر کلیدی برای برقراری پایداری در حل مسائل می‌باشد. گام زمانی به ترم نیروهای داخلی و خارجی، شرط سی اف ال و ترم انتشار لزجت بستگی دارد. به منظور اعمال شرایطی که ذرات، تحت تاثیر نیروهای داخلی و خارجی بیش از حد به همسایگان خود نزدیک نشوند داریم:

$$\delta t_f = \min \sqrt{\frac{h}{|f_a|}} \qquad (30)$$

که در رابطه بالا $h$ طول هموارسازی و $f_a$ نیروی داخلی یا خارجی بر واحد جرم است. ترکیب شرط سی اف ال و گام زمانی ویسکوزیته رابطه (31) را حاصل می کند.

$$\delta t_{cv} = \min_a \frac{h}{c_s + \max_b \left| \frac{h u_{ab} r_{ab}}{r_{ab}^2} \right|} \qquad (31)$$

در نهایت گام زمانی نهایی $\delta t$ به صورت زیر تعریف می شود که در آن عدد کورانت[3] می باشد.

$$\delta t = Cr \min(\delta t_f, \delta t_{cv}) \qquad (32)$$

عدد کورانت برای حفظ دقت حل عددی باید به اندازه کافی کوچک باشد. این عدد $1 \ge Cr > 0$ می باشد. در این تحقیق عدد کورانت برابر 0.2 استفاده شده است.

## 3- اعتبارسنجی عددی و شکست جت مایع

پس از بیان الگوریتم حل عددی به روش اس پی اچ ، قبل از شبیه‌سازی جریان شکست جت مایع، مساله نمونه جریان شکسته شدن سد با مانع برای اعتبار سنجی اولیه، شبیه سازی شده است.

### 3-1- جریان شکسته شدن سد با مانع

هندسه مساله شکست سد را می توان در شکل 2 مشاهده کرد. این مساله توسط اندازه عرض جریان سد $a$ بی بعدسازی شده است. ارتفاع سد به اندازه دو برابر عرض سد می باشد و همچنین مانع در مسیر جریان سد نیز با فاصله از سد قرار دارد. $d$ نیز بیانگر ابعاد مانع است. این مساله با استفاده از روش اس پی اچ تراکم ناپذیر توسط قاسمی [49] بررسی شده که از نتایج آن برای تایید صحت عملکرد آن استفاده شده است. لازم به توضیح است که نتایج قاسمی با نتایج موجود در مرجع [50] اعتبارسنجی شده است. در این شبیه سازی از آب با چگالی 1000 کیلوگرم بر مترمکعب به عنوان سیال استفاده و از فاز دوم (هوا) صرفنظر شده است. تعداد کل ذرات بکار گرفته شده برای این مساله 22803 ذره بوده است. فاصله اولیه ذرات برابر 0.002 متر، گام زمانی برابر 0.0001 ثانیه، سرعت اولیه صوت 17 متر بر ثانیه و ضریب لزجتی نیز 0.3 در نظر گرفته شده است. این مساله با بکارگیری دو تابع میانیاب شبیه سازی شده است. در شکل 3 می توان نتایج مدل سازی

---
1- Predictor-Corrector Scheme
2- Particle-based Methods
3- Courant Number

60

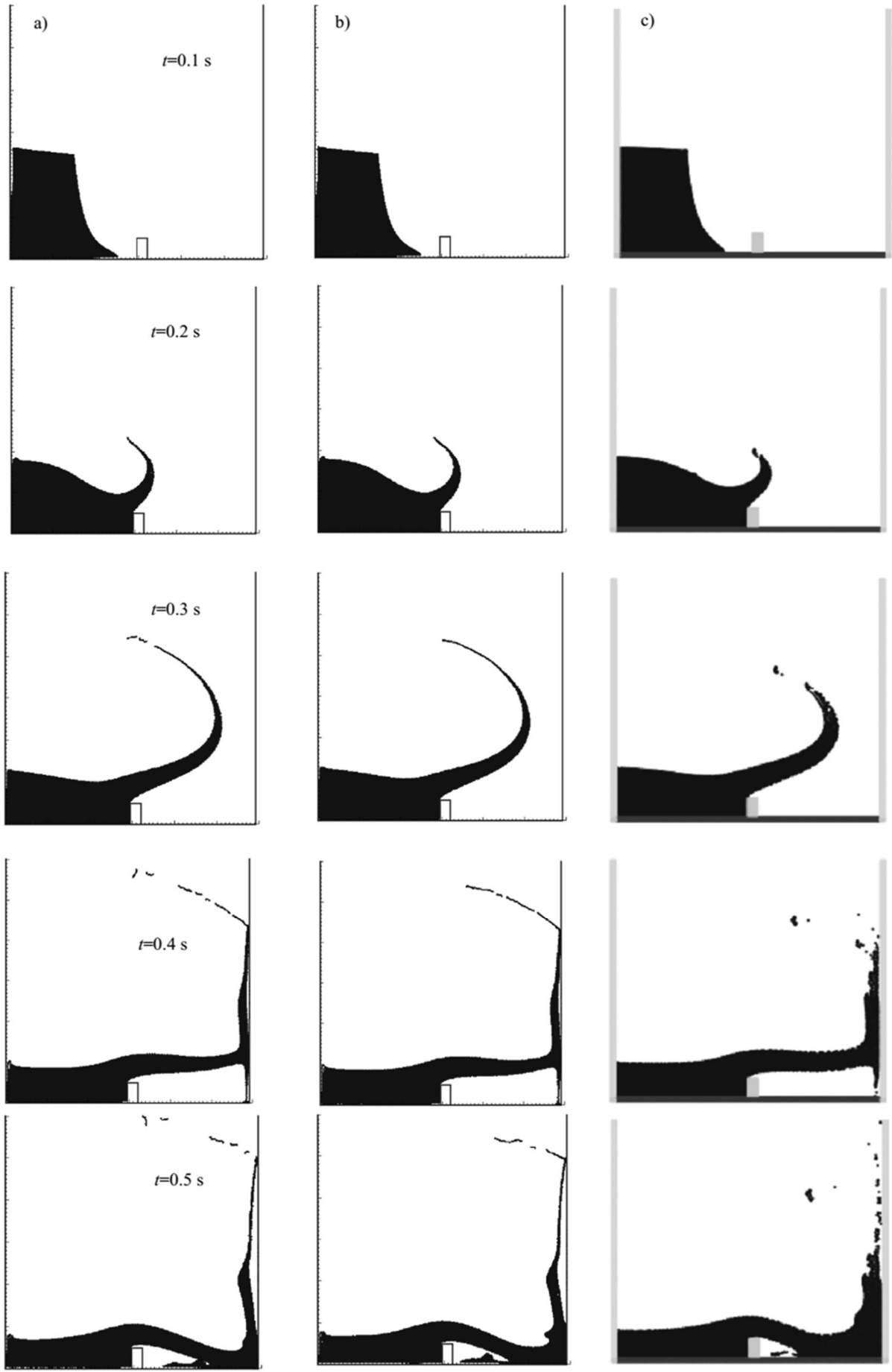

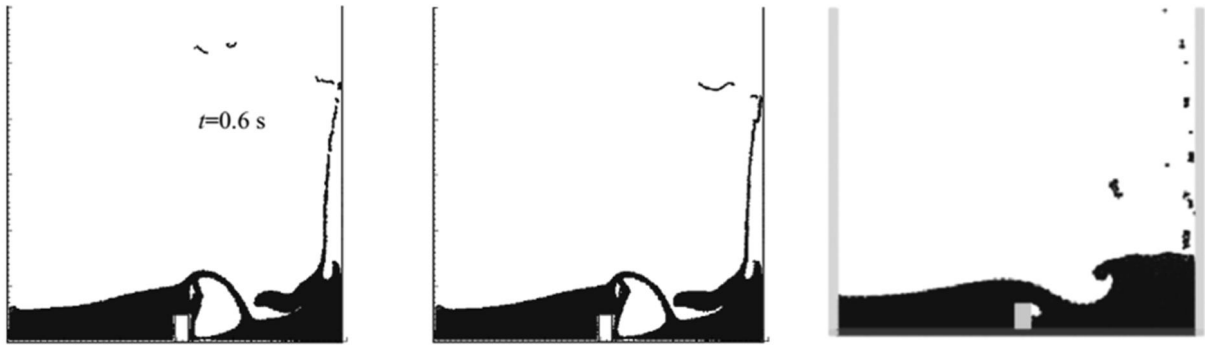

**Fig. 3** Comparison of dam-break with obstacle flow for 0.1 to 0.6 s. a) Present work for Wendland kernel b) Cubic spline kernel c) ISPH [49]

**شکل 3** مقایسه جریان شکست سد با مانع برای زمان های 0.1 تا 0.6 ثانیه. a: نتایج تحقیق حاضر برای تابع میانیاب وندلند، b: اسپلاین مکعبی، c: نتایج روش اس پی اچ تراکم ناپذیر [49]

میباشد. تمامی ابعاد با استفاده از قطر اریفیس بی بعدسازی شده است تا بتوان درک درست تری از ابعاد این نازل داشت. بنابراین نسبت منظری (ارتفاع نازل به قطر اریفیس) نازل در این تحقیق 15 می باشد که با توجه به بررسیهای انجام شده نسبت منظری بین 4 الی 20 در تحقیقات گذشته مرسوم بوده است. البته واضح است که می توان نسبت منظری نازل را برای وجود ذرات بیشتر افزایش داد اما این موضوع زمان محاسباتی را به شدت افزایش می دهد. در این مقاله شبیه‌سازی همانند مساله شکست سد با مانع برای دو تابع میانیاب اسپلاین مکعبی و وندلند انجام شده است و هدف اصلی به دست آوردن طول شکست جت تک فازی (صرفنظر کردن فاز هوا) از معادلات استاندارد روش اس پی اچ استفاده شده است. قطر اولیه نازل $D_o$= 0.01 m، لزجت دینامیکی آب $kgm^{-1}s^{-1}$ 0.001= $\mu$، $\sigma$ =72 mN/m، $\varphi_o$=1000 kg/m³، $g$=9.81 m/s²، سرعت اولیه صوت 10 متر بر ثانیه و ضریب لزجتی نیز 0.3 در نظر گرفته شده است. تعداد کل ذرات بکارگرفته شده برای این مساله 80175 ذره بوده است. روند جریان جت آب را شکل 5 شبیه سازی جریان جت تک فاز برای آب را نشان می دهد. جریان جت آب با سرعت اولیه $u_o$= 0.25 m/s با گذر زمان در این شکل دیده می شود.

این حالت از جریان جت به وسیله تابع میانیاب وندلند حل شده است. همچنین برای وضوح بیش تر، دو نمای بزرگ شده از لحظه‌ی شکست جت مایع نیز در قسمت پایین شکل 5 آورده شده است. با توجه به تعریف طول

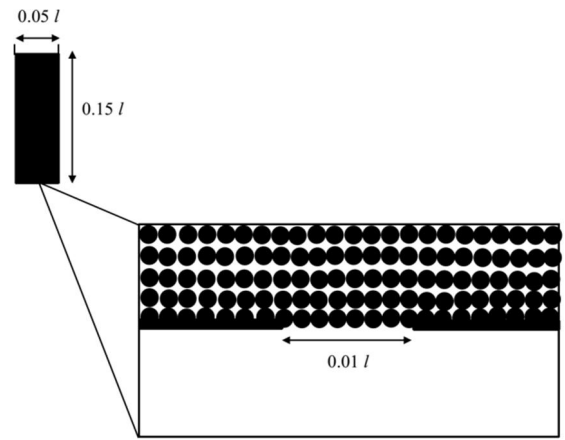

**Fig. 4** Jet geometry and solution domain for one-phase jet flow simulation in Rayleigh regime

**شکل 4** هندسه جت و دامنه حل برای شبیه سازی جت تک فازی در رژیم رِیلی

شکست جت مایع که عبارت از طول اولین نقطه‌ای که جریان پیوسته جت مایع از هم گسسته می‌شود، به تدریج قطر جت کم شده [19] و نهایتا در زمان جت $t$=0.78 s پیوسته از نقطه‌ای شکست میشود که در شکل 5 با عنوان طول شکست ($L$) مشخص شده است که مقدار این طول بر حسب قطر اریفیس بی بعد سازی شده است. همچنین همان طور که مشاهده می شود پس از خروج جت از نازل به دلیل اثرات کشش سطحی مایع، قطره در حال تشکیل است که اندازه آن حدود 1.98 برابر قطر اریفیس می‌باشد. براساس تئوری شکست رژیم ریلی، قطر قطره تشکیل شده حدود 1.89 برابر قطر اریفیس تخمین زده شده است که با شبیه سازی انجام شده در این تحقیق همخوانی دارد. به علت استفاده از شرط مرزی دافعه برای دیواره‌های نازل و در نتیجه اندرکنش ذرات مرزی و ذرات مایع در دهانه‌ی اریفیس، اندکی تقارن جریان جت در حل دو بعدی تحت تاثیر قرار گرفته است. همچنین نوسانات فشاری نیز در این عدم تقارن موثر هستند.

شبیه سازی‌های های چهار سرعت مختلف و برای دو تابع میانیاب انجام شده است. در مقایسه نتایج طول شکست به اعداد بی بعدی نیازمند هستیم که در ادامه این اعداد بی تعریف بی پرداخته شده است.

$$\mathbf{We} = \frac{\rho_o u_o^2 R}{\sigma} \tag{33}$$

$$\mathbf{Re} = \frac{\rho_o u_o R}{\mu} \tag{34}$$

$$\mathbf{Oh} = \frac{\sqrt{\mathbf{We}}}{\mathbf{Re}} \tag{35}$$

$$\mathbf{Fr} = \frac{u_o}{\sqrt{g D_o}} \tag{36}$$

روابط مذکور (33-36) به ترتیب بیانگر عدد وبر، رینولدز، اهنسرج[1] و فرود[2] هستند. همچنین در این روابط $R$ و $D_o$ شعاع و قطر اریفیس، $u_o$ سرعت اولیه جت، $\rho_o$ چگالی مایع، $\mu$ لزجت دینامیکی مایع، $\sigma$ ضریب کشش سطحی و $g$ شتاب گرانش می باشند. در رابطه (33) عدد وبر، We، بیانگر نسبت نیروی اینرسی بر کشش سطحی و در رابطه (34) عدد رینولدز، Re، برابر نسبت نیروی اینرسی بر نیروی لزجت است. نسبت عدد وبر و رینولدز، عدد بی بعد اهنسرج، Oh، را نشان می دهند (رابطه 35) که بیانگر نسبت نیروهای لزجتی بر کشش سطحی است. عدد فرود، Fr، در رابطه (36) نیز نسبت نیروهای اینرسی بر گرانشی را نشان می دهد. در جدول 1 مشخصات جریان و همچنین طول های شکست های به دست آمده به وسیله دو تابع میانیاب

---
1- Ohnesorge
2- Froude

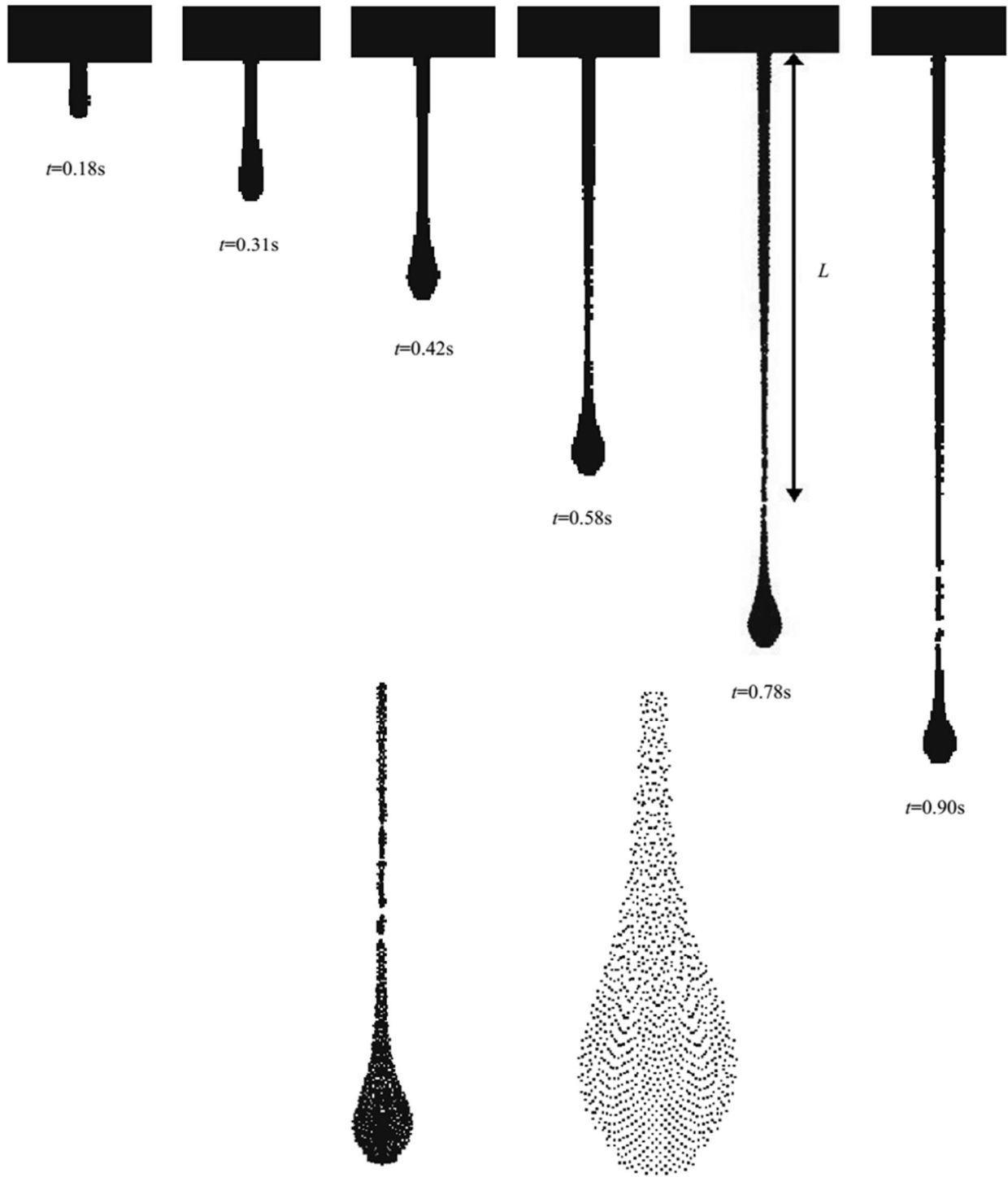

**Fig. 5** Up: Liquid jet flow evolution in time for uo=0.25 m/s. Breakup of liquid jet occurs in the instant t=0.78 s with L/d=28.10. Down: Fine resolutions of liquid jet breakup

**شکل 5** بالا: روند جریان جت مایع با گذر زمان برای $u_o$ =0.25 m/s. شکست جت مایع در زمان t = 0.78 s و L/d = 28.10. پایین: نماهای بزرگ شده از لحظه شکست جت مایع

اسپلاین و وندلند محاسبه شده است. برای اعتبار سنجی طول های شکست محاسبه شده، این طول ها با رابطه تجربی (37) که طول شکست بر قطر اریفیس بر اساس اعداد بی بعد می باشد [51]، مقایسه شده است که در ستون آخر جدول 1 مقادیر آن آورده شده است.

$$\frac{L}{D_o} = 13\sqrt{We}(1 + 30h) \qquad (37)$$

همانطور که در جدول 1 مشاهده می شود، طول شکست با استفاده از هر دو تابع میانیاب مقادیر قابل قبولی در مقایسه با رابطه تجربی داده اند اما تابع وندلند دارای مقادیر طول شکست نزدیکتری نسبت به رابطه تجربی دارد و در نتیجه خطای کمتری را در تمامی شبیه سازی های جریان جت منجر می شود. همچنین با توجه به عدد رینولدزها، جریان جت در رژیم جریان آرام قرار دارد. لازم به توضیح است که با توجه به معادله (10) مقدار لزجت برای

نقاط مختلف جریان متفاوت است اما این تغییرات کمتر از 5% بوده و می توان با تقریب خوبی عدد رینولدز را محاسبه کرد [19]. در این تحقیق هنگامی که فاصله دو ذره در امتداد جریان جت بیش تر از $1.5h$ شود، به عنوان معیار نقطه شکست جت مایع در نظر گرفته شده است. چرا که در این فاصله ذرات روی هم تاثیری نمی گذارند و ذرات وقتی این فاصله را از هم داشته باشند، دیگر مجددا به هم متصل نمی شوند بنابراین شکست رخ داده است. شکل های 6 و 7 به ترتیب تغییرات طول شکست جت مایع را بر حسب عدد رینولدز و عدد وبر برای دو تابع میانیاب اسپلاین و وندلند و همچنین رابطه تجربی (37) نشان می دهند. در این نمودارها طول شکست با افزایش عدد رینولدز و عدد وبر افزایش یافته است که با توجه به رژیم ریلی این قضیه روند درست نمودارها را نشان می دهد. همچنین در این دو نمودار مشهود است که تابع میانیاب وندلند طول شکست بلندتر و دقیقتری را نسبت به تابع میانیاب اسپلاین مکعبی محاسبه کرده است. در شکل 8 نتایج حاصل از روش اس پی اچ با روش ام پی اس [16] و نتایج تجربی [5] مقایسه شده است. همانطور که قبلا بیان شد، شیباتا و همکاران مسئله شکست جت مایع را با روش ام پی اس این شبیه سازی کرده اند اما از اثرات نیروی لزجتی صرفنظر کردهاند. بنابراین برای مقایسه نتایج روش اس پی اچ، در این حالت ما نیز با صرفنظر از لزجت، مسئله را دوباره حل کردیم. با توجه به شکل 8 و با توجه به اینکه مدل استفاده شده برای کشش سطحی برای دو روش عددی متفاوت بوده است، همخوانی خوبی بین دو روش عددی و همچنین نتایج تجربی وجود دارد.

## 4- جمع بندی و نتیجه گیری

در این تحقیق به شبیه سازی پدیده شکست جت مایع با استفاده از روش

هیدرودینامیک ذرات هموار پرداخته شد. بدین منظور، ابتدا معادلات حاکم بر جریان براساس روش اس پی اچ و با توسعه کد متن باز اسفیزیکس، الگوریتم بیان شده به وسیله مساله شکست سد با مانع مورد اعتبارسنجی قرار گرفت. در نهایت جریان جت مایع شبیه سازی و طول شکست مایع برای رژیم جریان ریلی در اعداد رینولدز کمتر از 1200 برای دو تابع میانیاب وندلند و اسپلاین مکعبی محاسبه شد. نتایج حاصله با استفاده از هر دو تابع میانیاب دارای دقت قابل قبولی بود، گرچه طول شکست بدست آمده با استفاده از تابع میانیاب وندلند دارای دقت بهتر و خطای کمتری نسبت به تابع میانیاب اسپلاین مکعبی در مقایسه با رابطه تجربی بود.

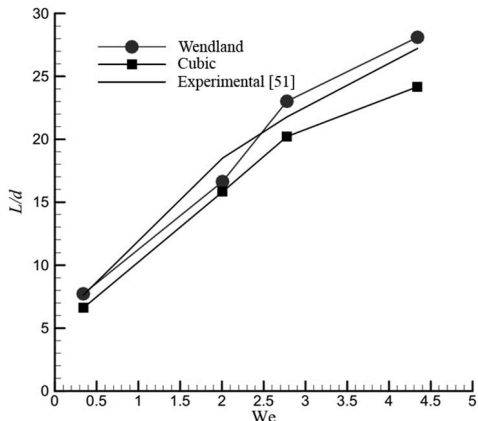

**Fig. 7** Variation of liquid jet breakup length with Weber number for spline and Wendland kernels in comparison to experimental correlation

**شکل 7** تغییرات طول شکست جت مایع بر حسب عدد وبر برای دو تابع میانیاب اسپلاین و وندلند در مقایسه با رابطه تجربی

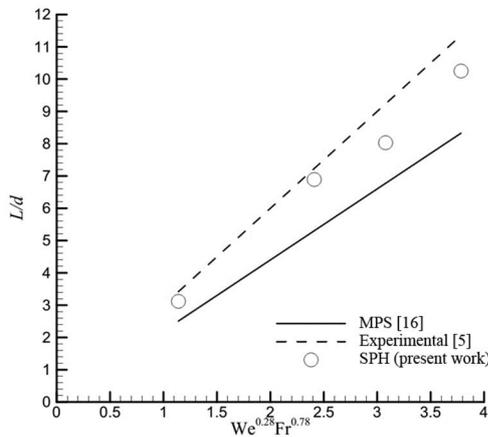

**Fig. 8** Comparison of SPH jet breakup lengths with MPS [16] and experimental results [5]

**شکل 8** مقایسه طول شکست جت با استفاده از روش اس پی اچ با نتایج ام پی اس و تجربی

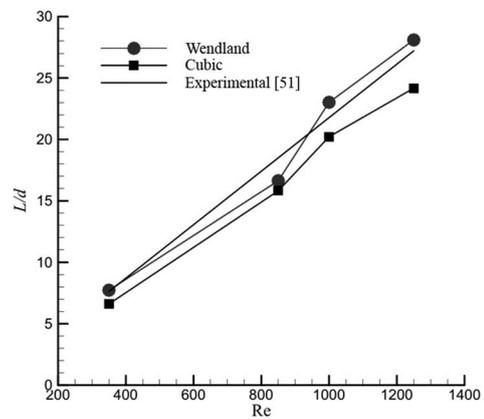

**Fig. 6** Variation of liquid jet breakup length with Reynolds number for spline and Wendland kernels in comparison to experimental correlation

**شکل 6** تغییرات طول شکست جت مایع بر حسب عدد رینولدز برای دو تابع میانیاب اسپلاین و وندلند در مقایسه با رابطه تجربی

**جدول 1** مشخصات جریان و طول شکست جت

**Table 1** Flow Characteristics and liquid jet breakup length

| خطای (%) وندلند | خطای (%) اسپلاین | $L/d$ [51] | $L/d$ وندلند | $L/d$ اسپلاین | Fr | We | Re | $u_o$ (m/s) | نوع جت |
|---|---|---|---|---|---|---|---|---|---|
| %1 | %13 | 7.62 | 7.73 | 6.61 | 0.22 | 0.34 | 350 | 0.07 | 1 |
| %10 | %14 | 18.51 | 16.64 | 15.83 | 0.54 | 2.01 | 850 | 0.17 | 2 |
| %6 | %7 | 21.78 | 23.02 | 20.22 | 0.64 | 2.78 | 1000 | 0.20 | 3 |
| %3 | %11 | 27.22 | 28.10 | 24.17 | 0.80 | 4.34 | 1200 | 0.25 | 4 |

## 5- فهرست علائم

| | |
|---|---|
| شتاب کشش سطحی ($ms^{-2}$) | a |
| ضریب ثابت معادله حالت تیت | B |
| سرعت صوت (m/s) | c |
| تابع رنگ | C |
| عدد کورانت | **Cr** |
| قطر نازل (m) | D |
| نیروی کشش سطحی بر واحد سطح (N) | f |
| نیروی کشش سطحی (N) | F |
| عدد فرود | **Fr** |
| شتاب گرانشی ($ms^{-2}$) | g |
| طول هموارسازی (m) | h |
| پارامتر بی بعد کننده ابعاد جت (m) | l |
| طول شکست مایع (m) | L |
| جرم (kg) | m |
| بردار عمود بر سطح تماس | n |
| عدد اهنسرج | **Oh** |
| فشار (Pa) | P |
| نسبت اندازه بردار مکان به طول هموارسازی | q |
| بردار مکانی هر ذره | q |
| شعاع نازل (m) | R |
| عدد رینولدز | **Re** |
| بردار سرعت ($ms^{-1}$) | u |
| عدد وبر | **We** |

### علائم یونانی

| | |
|---|---|
| ضریب لزجتی | α |
| ضریب ظرفیت گرمایی | γ |
| تابع نرمال کننده سطح | δ |
| انحنای سطح تماس | κ |
| لزجت دینامیکی ($kgm^{-1}s^{-1}$) | μ |
| لزجت مصنوعی | Π |
| چگالی ($kgm^{-3}$) | ρ |
| کشش سطحی | σ |
| گرادیان | ∇ |
| متغیر دلخواه | φ |
| ناحیه پشتیبانی تابع میانیابی | Ω |

### زیرنویس‌ها

| | |
|---|---|
| ذره دلخواه | a |
| ذره دلخواه | b |
| دو ذره دلخواه در اندرکنش | ab |
| ترکیب زمانی شرط کورانت و لزجتی | **cv** |
| نیروی داخلی یا خارجی | f |
| اولیه | o |
| سطح | s |
| لزجت | **vis** |

## 6- مراجع